\documentclass[dvips]{article}
\usepackage{icrctc07}

\newcommand{\td}{\textdegree}
\newcommand{\be}{\begin{enumerate}}
\newcommand{\ee}{\end{enumerate}}

\title{Observation of the Supernova Remnant IC 443 with VERITAS}
\authors{T. B. Humensky$^{1}$, for the VERITAS Collaboration$^{2}$.}
\shortauthors{T. B. Humensky and et al}
\afiliations{$^1$Enrico Fermi Institute, University of Chicago, 5640 S. Ellis Ave., 60637, Chicago, IL, USA\\ $^2$For full author list see G. Maier, "Status and Performance of VERITAS", these proceedings}
\email{humensky@uchicago.edu}

\abstract{Shell-type supernova remnants (SNRs) accelerate particles at the shock front between the expanding remnant and the swept-up interstellar medium. If these particles include protons and nuclei, very-high-energy gamma-ray emission may result from the decay of pions produced in interactions between cosmic rays and the local insterstellar medium. For SNRs that are interacting with a nearby molecular cloud, such as IC 443, the enhanced matter density provides a target medium that can amplify the gamma-ray emission. IC 443 also contains the pulsar wind nebula (PWN) CXOU J061705.3+222127. PWNe are the most plentiful galactic sources of very-high-energy gamma rays, which are produced in the shock formed at the collision of the pulsar wind with the ambient medium.

VERITAS is an array of four 12-m telescopes dedicated to gamma-ray astronomy in the energy band above 100 GeV. Located on Mt. Hopkins in southern Arizona, VERITAS operated during the 2006-2007 season in 2-, 3-, and 4-telescope observation modes. In this talk, results from three-telescope observations of the composite supernova remnant IC 443 during the 2006-2007 season are discussed.}

\begin{document}
\maketitle

\section{Introduction}
Fermi acceleration in supernova remnants (SNRs) has long been considered as one of the likely explanations for the origin of cosmic rays (CRs) up to the knee at $\sim 3\cdot 10^{15}\ \textrm{eV}$. This scenario drives the observations of SNRs in the very-high-energy (VHE) gamma-ray band. IC 443 is a prime candidate for such observations, due to its interaction with a nearby molecular cloud. Pion production, and the associated gamma-ray emission, is amplified by the cloud's high density (840 nucleon cm$^{-3}$) and total mass of $\sim 10^4\ \textrm{M}_\odot$~\cite{Torres2003a}.

IC 443 remains one of the best examples of an SNR interacting with molecular clouds. Its interaction with an H-II region to its northeast places IC 443 at a distance of $1.5\ \textrm{kpc}$~\cite{Fesen1984a}. From the central and southwest regions, OH maser emission indicates interaction with a molecular cloud~\cite{Claussen1997a} located along the line of sight. Petre \textit{et al.}~\cite{Petre1988a} find an age of $\sim 3000$ years for the shell based on modeling of soft X-ray emission; however, using a model of SNR/molecular cloud interactions, Chevalier estimates an age of $3\cdot 10^4$ years~\cite{Chevalier1999a}.

Near the southern edge of the shell, Chandra~\cite{Olbert2001a,Gaensler2006a} and XMM~\cite{Bocchino2001a} observations have resolved a pulsar wind nebula (PWN), J061705.3+222127, that may be associated with IC 443, though this association is not without its difficulties. Modeling of the PWN suggests that it is embedded in a hot, low-density medium, consistent with the shock-heated supernova ejecta rather than the interstellar medium or the molecular cloud. The temperature of the neutron star suggests an age of $\sim 3\cdot 10^4$ years~\cite{Gaensler2006a}, in agreement with the age of the shell estimated by~\cite{Chevalier1999a}. The direction of motion of the PWN does not point back to the center of the shell, though asymmetric expansion of the shell and distortion of the PWN tail by gas outflows could be exaggerating the apparent misalignment between the geometrical center of the SNR and the direction of motion of the PWN~\cite{Gaensler2006a}. If the PWN is not associated with IC 443, it may be associated instead with the overlapping SNR G189.6+3.3~\cite{Asaoka1994a,Olbert2001a}.

The EGRET source 3EG J0617+2238 is located at roughly the geometric center of the IC 443 shell, with a spectral index of -2.0 over the band $100\ \textrm{MeV - }30\ \textrm{GeV}$~\cite{Hartman1999a}. It is more likely that the gamma-ray emission is associated with the shocked molecular cloud, which it overlaps, than the PWN, which lies outside the 95\ \% confidence interval. At TeV energies, IC 443 was previously observed by several groups, including the Whipple collaboration, which set an upper limit on the flux of $6\cdot 10^{-12}\ \textrm{cm}^{-2}\textrm{s}^{-1}$ (0.11 Crab) above $500\ \textrm{GeV}$~\cite{Holder2005a}. The MAGIC collaboration has recently reported a detection of IC 443 at the level of 0.065 Crab above $100\ \textrm{GeV}$; the emission follows a power-law spectrum with a soft index of 3.1~\cite{Albert2007a}. 

VERITAS is an array of four $12\textrm{-m}$ telescopes located at an altitude of $1268\ \textrm{m}$ on Mt. Hopkins in southern Arizona, USA. Each telescope is equipped with a 500-pixel camera of $3.5\textrm{\td}$ total field of view. The array was completed in the spring of 2007. The energy threshold is roughly $150\ \textrm{GeV}$ for showers near zenith, with an energy resolution of $\sim 10\textrm{-}20\ \%$. The point-spread function, defined as the 68\ \% containment radius for gamma rays, is less than $0.14\textrm{\td}$. For a more complete description of VERITAS, see~\cite{Maier2007a}.

\section{Observations and Analysis}
VERITAS observed IC 443 with three telescopes during February and March of 2007, at an average zenith angle of $19\textrm{\td}$. The PWN location of RA 06 17 05.3, Dec +22 21 27 was tracked in wobble mode, in which the source is displaced by $0.5\textrm{\td}$ from the center of the field of view. Roughly equal amounts of data were acquired for each of four offset directions. The array trigger required two of three telescopes to trigger within $100\ \textrm{ns}$ in order to read out an event. After the application of quality-selection cuts on weather conditions and hardware and rate stability, and correction for system livetime, 15.9 hours of data remain. The data were analyzed following standard analysis procedures as discussed in~\cite{Cogan2007a,Daniel2007a}.

One systematic issue particular to this data set is the presence of two bright stars in the IC 443 field: HIP 29655 (V band magnitude 3.31), located $\sim0.53\textrm{\td}$ from the PWN; and HIP 30343 (magnitude 2.87), located $\sim1.37\textrm{\td}$ from the PWN. The stars add noise to nearby pixels and cause some pixels to have their HV turned off for a time when their PMT currents would be too high for safe operation. In the analysis, in order to recover some of the lost information, the suppressed pixels were assigned the average value of the signal in their unsuppressed neighbors (``pixel smoothing'').
\begin{figure}[h]
\begin{center}
\includegraphics [width=0.48\textwidth]{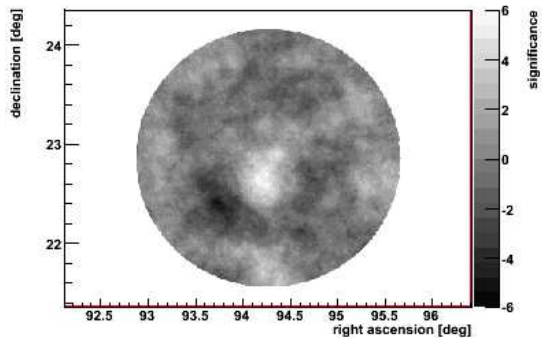}
\end{center}
\caption{Significance map for the IC 443 field, north wobble direction only. Note the deficit at (93.7, 22.5), the location of the star, and the excess at (94.2, 22.5), due to IC 443.}\label{fig:starstudy1}
\end{figure}
\begin{figure}[h!]
\begin{center}
\includegraphics [width=0.48\textwidth]{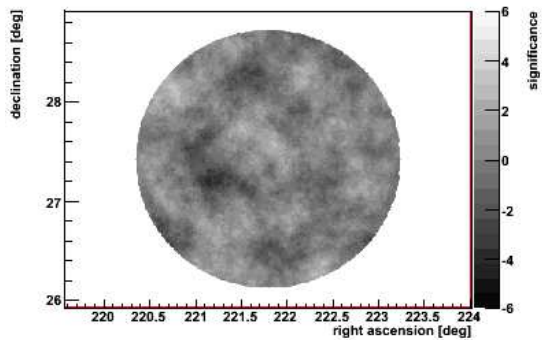}
\end{center}
\caption{Significance map of a bright-star field for which no gamma-ray sources are expected. Note as in Figure~\ref{fig:starstudy1} the deficit near the star location (221.2, 27.1); in this case no excess is visible.}\label{fig:starstudy2}
\end{figure}

In order to verify that the presence of a bright star in the field would not produce a false excess, observations were also made of a sky field that contained a magnitude-2.7 star (36 Eps Boo) but lacked any likely gamma-ray sources. Approximately 5 hours of data were taken in one wobble direction (called ``north''), the direction whose significance map showed the greatest deficit at the location of the star HIP 29655. Figure~\ref{fig:starstudy1} shows the significance map for the relevant set of IC 443 data, and Figure~\ref{fig:starstudy2} shows the map for the bright-star field: while both fields show a deficit at the location of the bright star, only the IC 443 field shows a significant excess. The same results are seen without the use of pixel smoothing; these and other tests (eg, rejecting camera images whose centroids lie close to a star, or applying a more stringent image-size cut) convince us that the effects of the bright stars are under control.

For this preliminary analysis, the event selection cuts were optimized for weak point sources using data taken on the Crab Nebula. Gamma/hadron separation was achieved using the so-called ``mean-scaled width'' and ``mean-scaled length'' parameters, as described in~\cite{Daniel2007a}. The ring background model was used for the background estimation since it is appropriate for potentially offset or slightly extended sources. The search radius was $0.158\textrm{\td}$.

\section{Results}
\begin{figure}
\begin{center}
\includegraphics [bb=0 36 696 490,clip,width=0.48\textwidth]{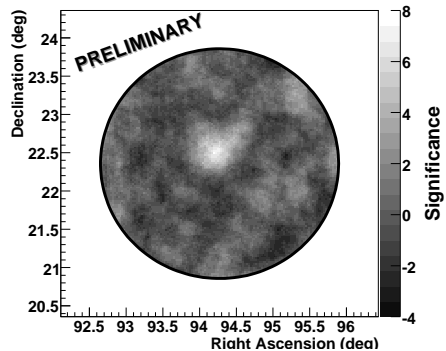}
\end{center}
\caption{Significance map for the IC 443 field.}\label{fig:sigmap}
\end{figure}

Figure~\ref{fig:sigmap} shows the significance map for the IC 443 field. The centroid of the excess, determined from a fit of a two-dimensional gaussian to an uncorrelated excess map, is located at 06 16 +22 30, consistent with the MAGIC position. The size of any possible systematic error in the centroid location is still under investigation. At the centroid location, an excess of 231 counts is found, with a significance of $7.1\ \sigma$ pre-trials and $6.0\ \sigma$ post-trials. The average gamma-ray rate was $0.24 \pm 0.04\ \gamma\textrm{/min}$, while the average background rate was $0.94 \pm 0.01\ \textrm{bg/min}$. The flux, determined by comparing the gamma-ray rate to that of the Crab Nebula at similar zenith angles ($11\textrm{\td}$), was roughly 3\textrm{-}4\ \% Crab above $200 \ \textrm{GeV}$, in agreement with that reported by MAGIC~\cite{Albert2007a}.

\begin{figure*}[th]
\begin{center}
\includegraphics [bb=0 36 696 582,clip,width=0.70\textwidth]{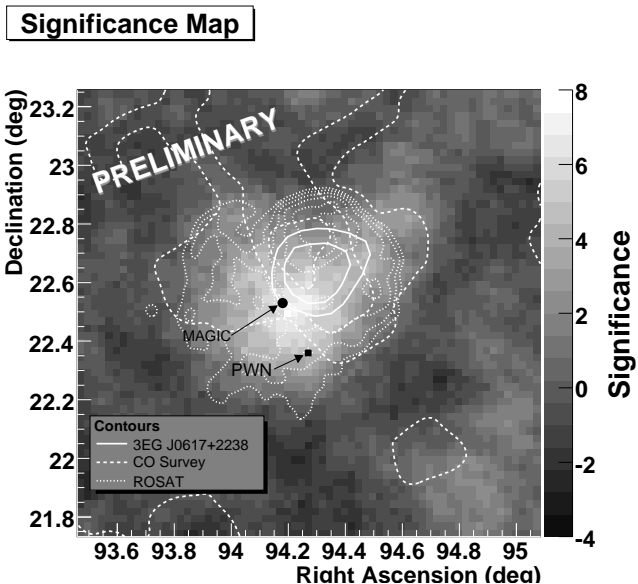}
\end{center}
\caption{Inner $1.5\textrm{\td}$ of the significance map for the IC 443 field, with multiwavelength features. Black circle: location of the centroid of the MAGIC detection~\cite{Albert2007a}. Black square: location of the PWN~\cite{Olbert2001a}. Solid white contours: 3EG J0617+2238~\cite{Hartman1999a}. Dotted white contours: ROSAT~\cite{Asaoka1994a}. Dashed white contours: CO survey~\cite{Huang1986a}.}\label{fig:mwlmap}
\end{figure*}

Figure~\ref{fig:mwlmap} shows the inner $1.5\textrm{\td}$ of the significance map. At the level of the sensitivity of this measurement, the TeV emission appears confined to a pointlike or near-pointlike region in the center of the remnant, rather than overlapping the radio/optical shell. The TeV emission is $\sim 0.15\textrm{\td}$ from the position of the PWN, which would seem to argue against an association. The spatial coincidence between the gamma-ray (TeV and GeV) emission, the densest region of the molecular cloud, and the maser emission is suggestive of hadronic cosmic-ray acceleration and interaction with the molecular cloud, which would provide a high density of target material for the production of VHE gamma rays~\cite{Torres2003a}.

\section{Conclusions}
VERITAS observations of the composite supernova remnant IC 443 in early 2007 with three telescopes have yielded a detection of a source with a flux of $\sim 3\ \%$ Crab above $\sim 200\ \textrm{GeV}$ at the level of $7.1\ \sigma$ pre-trials and $6.0\ \sigma$ post-trials. The preliminary flux and location are consistent with the MAGIC detection, and spectral and morphological analyses are in progress. While the nature of the TeV production remains to be determined conclusively, the coincidence of the gamma-ray emission with the dense cloud material and maser emission is suggestive of a hadronic origin.

\subsection*{Acknowledgments}

This research is supported by grants from the U.S. Department of Energy, the
 U.S. National Science Foundation, the Smithsonian Institution, by
NSERC in
Canada, by PPARC in the U.K. and by Science Foundation Ireland.

\bibliography{icrc1170}
\bibliographystyle{plain}

\end{document}